# The Social Psychology of Software Security (Psycurity)

Lucas Gren
Robert Feldt
lucas.gren@lucasgren.com
robert.feldt@chalmers.se
Chalmers | University of Gothenburg
Gothenburg, Sweden

## ABSTRACT

This position paper explores the intricate relationship between social psychology and secure software engineering, underscoring the vital role social psychology plays in the realm of engineering secure software systems. Beyond a mere technical endeavor, this paper contends that understanding and integrating social psychology principles into software processes are imperative for establishing robust and secure software systems. Recent studies in related fields show the importance of understanding the social psychology of other security domains. Finally, we identify critical gaps in software security research and present a set of research questions for incorporating more social psychology into software security research.

**ACM Reference Format:**
Lucas Gren and Robert Feldt. 2024. The Social Psychology of Software Security (Psycurity). In *Proceedings of International Workshop on Software Engineering in 2030 (SE 2030)*. ACM, New York, NY, USA, 5 pages. https://doi.org/10.1145/xxx.xxxx

## 1 INTRODUCTION

Software development in the contemporary landscape, particularly within agile teams, encounters multifaceted challenges driven by the unprecedented complexity and scale of modern systems. Beyond technical challenges, the sheer size of these systems necessitates collaboration from a large group of people. Previous research posits that the infusion of social psychology principles is essential to guide software processes (e.g., Gren et al. [9], Ralph et al. [19]), and in this paper, we show that it is also critical in the domain of software security. We argue that it is necessary to comprehend human behavior in software development organizations and that this behavior has a profound impact on software security, making a compelling case for the integration of social psychology studies to understand the dynamics of secure software engineering.

Before we look at the existing studies on psychology in software security, we would like to define psychology. According to Oxford Dictionary, *Psychology* is "*the scientific study of the human mind and its functions, especially those affecting behavior in a given context*" [1]. and *Social Psychology* is "*the branch of psychology that deals with social interactions, including their origins and their effects on the individual*" [2]. There are some studies in psychology on software security but those mainly focus on individual behavior. We argue that social interactions and their effect on the individual developer are essential to decrease vulnerabilities in code on a large scale.

The psychological research on people working in organizations (i.e., organizational psychology) is often divided into three different layers of abstraction: the individual (micro), the group (meso), and the organization (macro) [12]. An analogy we have often used when teaching psychology to software engineering students to explain these levels is that you cannot understand cancer by just studying atoms. To explain different phenomena in science the models need to be on the relevant abstraction level. Hence, we cannot understand the dynamics of secure software engineering without the meso and macro levels.

In contemplating the integration of social psychology studies into Software Engineering (SE), one might question their necessity. Should not results be universally consistent for individuals across the board? While true for some concepts, we would argue that the creation of complex software systems represents a novel human endeavor, introducing unique dynamics that challenge the assumption of uniformity in such outcomes.

As mentioned, the sheer scale of these software systems amplifies the need for collaborative efforts among individuals. Recognizing this, it becomes imperative to explore the applicability of social psychology findings in the context of software security. Findings in social psychology need to be tested in this new context, suggesting an uncharted territory with the potential for valuable insights.

Recently, the introduction of AI in the shape of Large Language Models has further highlighted the importance of understanding social dynamics in software engineering, including secure software development. Due to the ultra-fast development of AI in this context, and the trend of using AI agents to replace humans for certain tasks [30], AI Agents are soon intelligent partners in the development process.

AI agents, powered by LLMs and machine learning algorithms, are poised to become integral members of software engineering teams. These agents can soon act as:

- Code reviewers: AI agents can analyze code for style, efficiency, and potential security risks, supplementing human code reviews and improving overall code quality.
- Test automation assistants: AI agents can learn from existing test cases and generate new ones, streamlining the testing process and increasing security test coverage.
- Project managers: AI agents can track project progress, identify bottlenecks, and suggest solutions, alleviating the burden on human project managers and freeing time for organizational tasks.

The concept of a fully autonomous AI developer is still in the future. However, AI agents can significantly augment human capabilities, allowing engineers to focus on higher-level tasks requiring creativity, critical thinking, domain expertise, and a software security culture. We also need to understand this new social dynamic of people working with AI agents in the same team [25, 27], and how this affects secure software engineering.

## 2 CURRENT LANDSCAPE

### 2.1 Research on Software Security

Research in software security has primarily focused on technical methods to identify vulnerabilities in code [8]. The problem is that developers continue to produce code that is not secure, and the existing explanatory models are not sufficient [7, 31]. Another issue is that solutions often involve stricter controls, policies, or processes [4]. While these are often necessary, standards often contrast with the other development trends in software development that aim to make development more flexible in a rapidly changing environment with less focus on tools and processes (i.e., agile software development [15]). A proposed solution that has gained popularity is DevSecOps. However, a behavioral and cultural change within organizations is needed to integrate security thinking into DevOps, but it is still unknown what such a culture entails [23]. In a review article from 2022, the authors conclude that human factors are necessary to understand to implement DevSecOps, but there is not much research in this area [18]. Khan et al. [14] even concluded in 2021 that process-focused research on secure software development is a field of research that needs many more studies. All-in-all, there seems to be a general lack of studies that are not purely focused on the technical side of software security.

An examination of existing studies in software security reveals, we believe, a conspicuous lacuna in the exploration of social psychology aspects. While the industry has embraced fundamental technical and some process-oriented approaches as mentioned above, we argue that the psychological dimension is equally indispensable, if not more so, in ensuring comprehensive software security.

As mentioned, the two modern trends of more agility and software security compliance might seem contradictory [22]. However, we see a potential for synergy when security is deeply ingrained within the organization's culture and team norms.

Agile principles emphasize responsiveness to change [10], but these could be seamlessly integrated with a focus on security. When core human values prioritize security, as reflected in established group norms and the overall organizational culture, secure practices become a natural part of the workflow. Organizational culture encompasses ingrained operational habits and the values upheld within the company [26].

In this scenario, compliance with security standards becomes a final quality checkpoint, validating existing secure practices embedded within the company's ethos. Compliance does not add additional burdens; it simply formalizes existing practices, minimizing the need for further effort.

### 2.2 Studies on Security in Related Fields

In the field of information security, a meta-study from 2019 [6] concluded that compliance with information security policy highly depends on psychological aspects like personal norms and ethics, attitudes, and normative beliefs (all of which are explained in Section 3), and much more so than on response cost or threat severity. The security of IoT technologies has also been shown to depend on human factors to a large extent [5], which is also well-known in research on social engineering in hacker attacks [28]. Together, these studies in related fields show that it is highly likely that social psychology aspects have a large influence on secure software engineering.

### 2.3 Psychology and Software Security

We only found one study looking at individual psychology within secure software development [16]. That study examines what motivates developers to write secure code. The paper concludes that an individual must have the capability, motivation, and opportunity to display a certain behavior. The paper is a great first step but we need more studies on these different aspects.

The only study we found on social psychology in software security also laments the persistence of developers writing insecure code, despite extensive efforts within the security community [21]. Even if the study also claims to apply social psychology theories, the authors urge a more nuanced understanding of psychological factors, marrying technical sophistication with psychological frameworks and usability. Acknowledging the cognitive processes at play, the study emphasizes the underexplored influence of social psychology literature [21]. These same authors seem to be the only ones having teamed up with social psychologists to explore more of the social psychology of software security [20], which we predict many more researchers need to do shortly and for years to come. If not, society will be at risk since we will not be able to build secure software systems at the level society needs.

## 3 PROPOSED RESEARCH DIRECTIONS

Applying psychology to secure software engineering has a wide range of possibilities due to the few studies conducted so far. A key to what is missing, in general, are studies on the meso and macro levels with the notable exception of Rauf et al. [20]. Below we describe two general strains of research needed that apply social and organizational psychology with a set of research questions, but many others need to be explored.

### 3.1 Proposition 1: Group Norms

We propose the application of group norms in the domain of secure software engineering. Hogg and Vaughan [13] define norms as *"attitudinal and behavioral uniformities that define group membership and differentiate between groups."* One theory from management science that would be relevant is the concept of mindful organizing for safety (MO) in high-reliability organizations [29]. Mindful organizing for safety advocates that many organizations could benefit from emulating high-reliability organizations [17]. We believe secure software development would greatly benefit from such an emulation. It is important to note that mindfulness in this context refers to retrospection, not the contemporary psychological mindfulness-based intervention movement based on Buddhist traditions [24].

Martínez-Córcoles and Vogus [17] outline the principles of MO, encompassing actions and interactions through which teams anticipate, prevent, and dynamically respond to errors and unexpected events.

These principles include:

(1) Preoccupation with Failure: MO encourages routine discussions that explore potential failure scenarios. This fosters the identification of early warning signs and facilitates preventative action to mitigate future problems.

(2) Reluctance to Simplify Interpretations: MO promotes a culture of questioning existing procedures and protocols. Teams are encouraged to engage in critical evaluation to identify potentially more reliable alternatives, fostering continuous improvement.

(3) Integrated Understanding of Operations: MO emphasizes the creation and maintenance of a shared mental model within the team regarding operational processes. This ensures all members possess a comprehensive understanding of how different aspects of the organization interrelate.

(4) Commitment to Resilience: MO underscores the importance of learning from setbacks. Through thorough analysis and discussion of encountered issues, teams can foster resilience and improve their ability to recover effectively from unexpected events.

(5) Expertise Over Authority: MO emphasizes the prioritization of expertise when resolving problems. This principle encourages teams to value knowledge and experience above hierarchical rank, ensuring the optimal solution is implemented [17].

MO is not confined to a single level within an organization but spans across the various abstraction levels described before, from top management to front-line employees. The research emphasizes that true MO is a cross-level phenomenon that requires coordination and interaction among various organizational abstraction levels. The article specifically notes that while certain aspects of MO, like preoccupation with failure and sensitivity to operations, are present across all levels, some processes, like the reluctance to simplify interpretations and commitment to resilience, are more prevalent at the macro level [17]. This highlights the importance of tailored approaches to MO that consider the unique roles and responsibilities at each level, which then need to be adapted to secure software engineering.

Values play a crucial role in MO by aligning attitudes and behaviors within the organization. MO emphasizes the importance of not just valuing safety and reliability but also disvaluing mis-specifications, mis-estimations, and misunderstandings. Martínez-Córcoles and Vogus [17] point out that organizations with high levels of MO often embody values that promote careful attention to detail and a continuous questioning attitude that prevents errors and enhances safety. This approach encourages a culture where organizational values support the vigilant and proactive behaviors that are central to MO [17].

Leadership is critical in fostering and sustaining MO within an organization. Martínez-Córcoles and Vogus [17] discuss empowering leadership, which includes behaviors such as modeling, coaching, participative decision-making, and showing concern for others' welfare. This type of leadership supports MO by encouraging open communication, collaborative problem-solving, and the distribution of decision-making authority (cf. agile leadership [11]). Martínez-Córcoles and Vogus [17] suggest that leaders who adopt an empowering style can effectively instill MO practices by enabling employees to take an active role in safety and reliability processes.

External stakeholders, including regulators and other outside bodies, play a significant role in shaping MO within organizations. Martínez-Córcoles and Vogus [17] illustrate that engaging with external stakeholders can enhance MO by keeping safety rules and regulations salient and up-to-date. Such engagement helps organizations remain vigilant and responsive to emerging safety concerns and regulatory requirements. This reciprocal relationship between organizations and regulators not only fosters a mindful approach within the organization but also allows organizations to influence the broader regulatory environment to support MO practices. An increasing number of software companies need to adhere to new regulations, e.g., the NIS-2 directive in Europe[1].

Since this framework has not been applied to software security, all of these aspects warrant empirical research. We are, though, aware that many software security-focused companies do apply all or a subset of these practices, however, research lags behind. Below are ten research questions that guide how high-security norms can be understood through the lens of mindful organizing.

---

[1] https://digital-strategy.ec.europa.eu/en/policies/nis2-directive

**MO-RQ1:** How can mindful organizing principles be adapted to enhance cybersecurity resilience in software development teams?

**MO-RQ2:** What role do empowering leadership styles play in fostering a culture of mindful organizing within software development teams, and how does this impact software security?

**MO-RQ3:** How does the cross-level and cross-unit distribution of mindful organizing influence secure software development practices across different teams and departments?

**MO-RQ4:** What are the specific human resource practices that can support the development and sustenance of mindful organizing in software development environments, particularly those dealing with security?

**MO-RQ5:** How can external stakeholders, such as regulatory bodies or cybersecurity firms, influence the adoption and effectiveness of mindful organizing practices in secure software development?

**MO-RQ6:** In what ways can the principles of mindful organizing be used to develop and refine security protocols and frameworks for software development?

**MO-RQ7:** What measures and metrics can be used to assess the impact of mindful organizing on the reduction of security vulnerabilities and incidents in software development?

**MO-RQ8:** How do specific training and mentoring approaches contribute to the cultivation of a mindful organizing mindset among software developers, particularly in handling security issues?

**MO-RQ9:** Can the implementation of mindful organizing practices improve the ability of software development teams to anticipate, prevent, and respond dynamically to cybersecurity threats?

**MO-RQ10:** What are the challenges and barriers to embedding mindful organizing in the software development lifecycle, particularly in phases critical to security, and how can they be overcome?

These questions aim to bridge the gap between theoretical principles of mindful organizing and their practical application in the realm of secure software development.

### 3.2 Proposition 2: Social Forces Driving Behavior

To understand the intricate web of social forces influencing behavior in secure software engineering, we turn to the Theory of Planned Behavior (or TPB) [3]. This well-established psychological framework posits that behavior is driven by three main factors: (1) attitude toward the behavior, (2) subjective norms, and (3) perceived behavioral control.

In the context of secure software development, we need to investigate how developers' attitudes towards security practices, the influence of social norms within development teams, and the perceived control over implementing secure coding practices collectively shape behavior. By unraveling these psychological dynamics, we anticipate developing more effective interventions tailored to the specific motivations and intentions guiding developers' actions.

*Attitude Towards the Behavior.* **TPB-RQ1:** How do developer attitudes towards the benefits and drawbacks of secure coding practices (e.g., threat modeling or code review) influence their intention to adopt those practices?

**TPB-RQ2:** Does a positive attitude towards the effectiveness of secure coding practices in preventing vulnerabilities lead to a stronger intention to implement them?

**TPB-RQ3:** Can negative perceptions regarding the time or complexity associated with secure coding practices create a barrier to their adoption?

These aspects are partially investigated in Larios-Vargas et al. [16].

*Subjective Norms.* **TPB-RQ4:** How do the expectations and pressures from team leads, security champions, and peers (subjective norms) influence a developer's intention to adopt secure coding practices?

**TPB-RQ5:** Does a team culture that emphasizes secure coding practices create a stronger subjective norm, leading to increased adoption among developers?

**TPB-RQ6:** How does the perceived pressure to meet deadlines or deliver features impact a developer's intention to prioritize secure coding practices?

*Perceived Behavioral Control.* **TPB-RQ7:** Does a developer's perception of their own skills and knowledge regarding secure coding practices influence their intention to adopt them?

**TPB-RQ8** How does the availability of security tools, training opportunities, and clear coding standards within the development environment impact a developer's perceived control over implementing secure coding practices?

**TPB-RQ9:** Can a lack of confidence in one's ability to perform secure coding effectively act as a barrier to their adoption, even with a positive attitude and supportive social norms?

These research questions explore the theoretical relationships between the TPB constructs and developer intention towards secure coding practices. By investigating these relationships, the research can provide valuable insights into how to promote the adoption of secure coding practices within software development teams.

## 4 CONCLUSION

In summation, this visionary exploration advocates for the integration of social psychology into the fabric of software security practices. Beyond the binary realm of technical proficiency and process adherence, we underscore the nuanced human factors that significantly influence the security posture of software systems. By addressing individual, group, and organizational dimensions, we posit that a richer understanding of human behavior can usher in a new era of secure software engineering.